\documentclass[preprint]{elsarticle}

\usepackage{euscript,amssymb,amsmath}

\usepackage{graphicx}
\usepackage{epsfig}

\newcommand{\be}[1]{\begin{equation}\label{#1}}
\newcommand{\ee}{\end{equation}}
\newcommand{\ba}[1]{\begin{eqnarray}\label{#1}}
\newcommand{\ea}{\end{eqnarray}}
\newcommand{\rf}[1]{(\ref{#1})}

\begin{document}

\begin{frontmatter}

\title{Remarks on gravitational interaction\\ in Kaluza-Klein models}

\author{Maxim Eingorn}
\ead{maxim.eingorn@gmail.com}

\author{Alexander Zhuk}
\ead{ai.zhuk2@gmail.com}

\address{Astronomical Observatory and Department of Theoretical Physics,\\ Odessa National University, Street Dvoryanskaya 2, Odessa 65082, Ukraine}

\begin{abstract}
In these remarks, we clarify the problematic aspects of gravitational interaction in a weak-field limit of Kaluza-Klein models. We
explain why some models meet the classical gravitational tests, while others do not. We show that variation of the total volume of the internal spaces generates the
fifth force. This is the main reason of the problem. It happens for all considered models (linear with respect to the scalar curvature and nonlinear $f(R)$, with
toroidal and spherical compactifications). We explicitly single out the contribution of the fifth force to nonrelativistic gravitational potentials. In the case of
models with toroidal compactification, we demonstrate how tension (with and without effects of nonlinearity) of the gravitating source can fix the total volume of the
internal space, resulting in the vanishing fifth force and consequently in agreement with the observations. It takes place for latent solitons, black strings and black
branes. We also demonstrate a particular example where non-vanishing variations of the internal space volume do not contradict the gravitational experiments. In the case
of spherical compactification, the fifth force is replaced by the Yukawa interaction for models with the stabilized internal space. For large Yukawa masses, the effect
of this interaction is negligibly small, and considered models satisfy the gravitational tests at the same level of accuracy as general relativity.
\end{abstract}


\end{frontmatter}


\section{\label{sec:1}Introduction}

\setcounter{equation}{0}

In our recent papers \cite{EZ3}-\cite{ChEZ2}, we investigated classical gravitational tests (the perihelion shift, the deflection of light and the time delay of radar
echoes) in Kaluza-Klein (KK) models. We paid attention mainly to models with toroidal compactification of extra dimensions, i.e. with compact and flat internal spaces.
On the one hand, these theories are very popular in the literature devoted to KK models. On the other hand, it is well known that the gravitational tests in general
relativity are calculated on the background of a flat metrics. This background metrics is perturbed by a pointlike mass. In a weak-field limit, such matter source has a
dustlike equation of state. In general relativity, this formulation of the problem leads to formulas for gravitational tests, which are in good agreement with the
experimental data \cite{Landau}. Therefore, to generalize this approach, we also supposed that the background metrics (when the matter source is absent) is flat for our
external four-dimensional spacetime and internal spaces, and a pointlike matter source has a dustlike equation of state in all spatial dimensions. To our surprise, this
approach does not work in KK models. Obtained formulas strongly contradict the observations \cite{EZ3}. It is important to note that the result does not depend on the
pointlike\footnote{In the most of cases, when we use the word "pointlike", we usually mean a gravitating mass which has a deltashaped form in the external space and is
uniformly smeared over the internal space. In this case, the nonrelativistic gravitational potential exactly coincides with the Newtonian one \cite{EZ2}.} approximation.
Instead of the deltashaped form, we can consider a compact object in the form of a perfect fluid with the dustlike equation of state in all spatial dimensions, and we
obtain the same negative result \cite{EZ5}. It turned out that to satisfy the experimental data, the matter source should have negative equations of state (tension) in
the internal spaces \cite{EZ4,EZ5}. For example, latent solitons have such tension and they satisfy the gravitational tests at the same level of accuracy as general
relativity \cite{EZ5}. The uniform black strings and black branes are particular examples of the latent solitons. The similar situation takes place for nonlinear (with
respect to the scalar curvature) KK theories with toroidal compactification \cite{EZf(R),EZf(R)solitons}. Here, a pointlike mass with the dustlike equation of state in
all spatial dimensions contradicts the observations \cite{EZf(R)}, but there are two classes of asymptotic latent solitons, which are in agreement with the observations
at the same level of accuracy as general relativity \cite{EZf(R)solitons}. For both of these classes, a gravitating mass has tension in the internal space.

For black strings and black branes, the notion of tension is defined, e.g., in \cite{TF} and it follows from the first law for black hole spacetimes \cite{TZ,HO,TK}. We
study gravitational experiments for KK models in the Solar system. Obviously, our Sun is not a relativistic astrophysical object. Therefore, to calculate the motion of a
test body in the vicinity of the Sun, there is no need to take into account the relativistic properties of black holes/strings (e.g., the presence of the horizon, a
possible phase transition between black strings and black branes \cite{Kol}, etc). It is sufficient to use  the corresponding black strings/branes metrics in the
weak-field limit. Additionally, it is worth noting that the metric coefficients for uniform black strings/branes depend only on the three-dimensional radius-vector.
Therefore, a matter source is uniformly smeared over the extra dimensions and the nonrelativistic gravitational potential exactly coincides with the Newtonian one.
Strictly speaking, to be in agreement with gravitational experiments, it is not sufficient to have the "correct" nonrelativistic gravitational potential. All models with
the "smeared" extra dimensions have the Newtonian potential in the weak-field limit (see, e.g., the class of the exact solitonic solutions \cite{EZ5}). To achieve the
concordance with the experimental data, the temporal and three-dimensional spatial metric coefficients should relate correspondingly with each other. The uniform black
strings/branes have the right relation but the sources with the dustlike equation of state in all spatial dimensions do not have.

Then, we generalize our studies to the case of KK models with spherical compactification of the internal space \cite{ChEZ1,ChEZ2}. Here, the background metrics is not
flat because the internal space (e.g., the two-sphere) is curved. To create such background, we need to introduce a background matter source, e.g., in the form of a
perfect fluid. A pointlike mass disturbs this background. We demonstrate that there are two different models with and without a bare cosmological constant. In the former
case, a pointlike mass contradicts the observations \cite{ChEZ1}. However, in the latter case, the perturbed metric coefficients have the Yukawa type corrections with
respect to the usual Newtonian gravitational potential \cite{ChEZ2}. These corrections are negligible in the Solar system. Therefore, the considered model satisfies the
gravitational tests.

The main objective of this paper is to explain why some models with toroidal and spherical compactifications fail with the observations, while the others are in good
agreement with the experimental data. It is generally accepted that the variations of the internal space volume produce the fifth force which may lead to the
contradiction with the observations. Therefore, first, we single out explicitly the contribution of such variations to nonrelativistic gravitational potentials. We show
that the variation of the total volume of the internal spaces generates the fifth force for the models which fail with the observations. This is the main reason of the
problem in these models. Then, we demonstrate that tension can fix the total volume of the internal space, eliminating the fifth force. Second, we ask the following
question. Is a vanishing variation of the internal space volume a necessary and sufficient condition of the agreement with the gravitational tests? To our surprise, the
answer is negative. We find a particular example where non-vanishing variations of the internal space volume do not lead to the contradiction with the gravitational
experiments. Moreover, in this model, we can make the variations equal to zero, but then the gravitational tests fail here. We also consider briefly the model with a
pointlike gravitating source in the case of spherical compactification of the internal space. Here, tension is absent and the fifth force is replaced by the Yukawa
interaction. For sufficiently large Yukawa masses (it happens, e.g., in the Solar system), the effect of this interaction is negligibly small, and the considered model
satisfies the gravitational tests at the same level of accuracy as general relativity.

The paper is structured as follows. In Sec. 2, we consider linear models with toroidal compactification of the internal spaces and different types of gravitating
sources. These models are generalized to nonlinear $f(R)$ ones in Sec. 3. In Sec. 4, we investigate models with spherical compactification of the internal space. The
main results are summarized in the concluding Sec. 5. In appendix, we demonstrate that effective four-dimensional approach confirms the results obtained in Sec. 2.

\section{Linear models with toroidal compactification}

In this section we analyze linear with respect to the scalar curvature KK models with toroidal compactification of the internal spaces.

\subsection{Pointlike mass\label{1.1s}}

First, we investigate a model with a pointlike massive source. We consider a weak-field limit. It means that the gravitational field is weak, i.e. the metrics is only
slightly perturbed from its flat spacetime value:
\be{1}
g_{ik}\approx \eta_{ik}+h_{ik}\, .
\ee
Here, the metric perturbations $h_{ik} \sim O(1/c^2)$, where  $c$ is the speed of light, $i,k =0,1,\ldots ,D$, and $D$ is a total number of the spatial dimensions. In
the weak-field limit, the only nonzero component of the energy-momentum tensor for a pointlike mass at rest is $T_{00} \approx \rho_D c^2 \sim O(c^2)$.
$\rho_D$ is a $D$-dimensional rest mass density, and for a pointlike mass $m$ we have $\rho_D =m\delta ({\bf r}_D) $. Usually, we deal with the case of matter sources,
which are uniformly smeared over the extra dimensions \cite{EZ2}. In this case, the metric coefficients may depend only on coordinates of the external space\footnote{For
the smeared extra dimensions, KK modes are absent. Our following analysis can be easily generalized to the case of non-smeared extra dimensions. In this case, KK modes
are present in considered models. However, in the Solar system, they are negligible compared with the zero mode \cite{EZ3}, and such generalization does not change the
main conclusions of our paper.}. For the smeared extra dimensions, the nonrelativistic three-dimensional mass density $\rho_3$ is connected with the $D$-dimensional one
as follows: $\rho_D = \rho_3/V_{D'}=m\delta ({\bf r}_3)/V_{D'}$, where $D'=D-3$ is a total number of the extra dimensions and $V_{D'}$ is a total volume of the
unperturbed internal spaces. For example, if $a_i$ are periods of tori, then $V_{D'}=\prod_{i=1}^{D'} a_i$.  For such setup, the Einstein equation
\be{2}
R_{ik}=\frac{2S_D\tilde G_{\mathcal{D}}}{c^4}\left(T_{ik}-\frac{1}{D-1}g_{ik}T\right)\, , \ee
where $S_D=2\pi^{D/2}/\Gamma (D/2)$ is a total solid angle (a surface area of the $(D-1)$-dimensional sphere of the unit radius) and $\tilde G_{\mathcal{D}}$ is the
gravitational constant in the $(\mathcal{D}=D+1)$-dimensional spacetime, is reduced to a system of linearized equations with the following nonzero solutions
\cite{EZ3}:
\ba{3}
h_{00}&=&-\frac{2(D-2)}{D-1}\frac{2G_N m}{c^2 r_3}\, ,\\
h_{\alpha\alpha}&=&-\frac{2}{D-1}\frac{2G_N m}{c^2 r_3}\, ,\quad \alpha = 1,2,3\, ,\label{4}\\
h_{\mu\mu}&=&-\frac{2}{D-1}\frac{2G_N m}{c^2 r_3}\, ,\quad \mu = 4,5,\ldots ,D\, ,\label{5}
\ea
where we introduce the Newton's gravitational constant
\be{6}
G_N=\frac{S_D}{4\pi} \frac{\tilde G_{\mathcal{D}}}{V_{D'}}\, .
\ee
Hereafter, the Latin indices $i,k = 0,\ldots,D$, the Greek indices $\alpha,\beta = 1,2,3$ and the Greek indices $\mu,\nu = 4,5,\ldots ,D$.

It is well known that to satisfy the gravitational experiments (the deflection of light, the time delay of radar echoes) at the same level of accuracy as general
relativity, the metric coefficients $h_{00}$ and $h_{\alpha\alpha}$ should coincide with each other. However, Eqs. \rf{3} and \rf{4} show that for the considered model
$h_{00}/h_{\alpha\alpha}=D-2$ and this ratio does not depend on the size of the internal space. So, we can not make it equal to unity. On the other hand, $h_{00}$
defines the nonrelativistic gravitational potential: $h_{00}=2\varphi/c^2$.  For example, in general relativity $\tilde h_{00}=2\varphi_N/c^2 = -2G_N m/\left(c^2
r_3\right)$ and $\tilde h_{00} =\tilde h_{\alpha\alpha}$. In our case, the Newtonian gravitational potential acquires a prefactor $2(D-2)/(D-1)$. What is the reason for
this prefactor and for inequality between $h_{00}$ and $h_{\alpha\alpha}$? It can be easily seen that we can rewrite $h_{00}$ and $h_{\alpha\alpha}$ in the following
form:
\ba{7}
h_{00} &=& \tilde h_{00} +\frac12 \delta V_{D'}\, ,\\
h_{\alpha\alpha} &=& \tilde h_{\alpha\alpha} -\frac12 \delta V_{D'}\, ,\label{8}\
\ea
where $\delta V_{D'}$ is a relative value of the internal space volume variation:
\be{9}
\delta V_{D'} =\sum_{\mu=4}^D h_{\mu\mu}= -\frac{2(D-3)}{D-1}\frac{2G_N m}{c^2 r_3}\, .
\ee
Eqs. \rf{7} and \rf{8} demonstrate explicitly that the prefactor $2(D-2)/(D-1)$ in $h_{00}$ as well as inequality between $h_{00}$ and $h_{\alpha\alpha}$ originate from
the admixture of the variation of the internal space volume. Comparing $\delta V_{D'}$ with $2\varphi_N/c^2$ shows that it coincides with the Newtonian expression up to
the prefactor $2(D-3)/(D-1)$ and satisfies the equation similar to the Poisson one. Obviously, this leads to the additional force which is proportional to the Newtonian
expression. Usually such force is called the fifth force. Therefore, in this model, the variation of the internal space volume results in the fifth force. This is the
reason of the contradictions with the experimental data.

\subsection{Latent solitons, black strings and black branes\label{1.2s}}

Above, we consider the case of a pointlike mass with the dustlike equations of state in all spatial dimensions. However, there is a class of exact soliton solutions
(see, e.g., \cite{EZ4,EZ5}) with nonzero equations of state in the extra dimensions:
\be{10} T_{\mu\mu}\approx \omega_{\mu}\frac{1}{V_{D'}}\rho_3({\bf r}_3)c^2\approx\omega_{\mu}T_{00},\quad \mu=4,5,\ldots,D\, . \ee
These solutions are defined by the parameters $\gamma_{\mu}$, which are connected with the equation of state parameters $\omega_{\mu}$ \cite{EZ5}:
\be{11} \omega_{\mu}=\frac{\gamma_{\mu}-1+\tau}{2-\tau}\, , \ee
where $\tau = \sum_{\mu=4}^D \gamma_{\mu}$.

In the weak-field limit, the metric perturbations for these solutions read \cite{EZ5}:
\ba{12}
h_{00}&=&-\frac{2(D-2)}{D-1}\frac{2G_N m}{c^2 r_3}
-\frac{2\Omega}{D-1}\frac{2G_N m}{c^2 r_3}\, ,\\
h_{\alpha\alpha}&=&-\frac{2}{D-1}\frac{2G_N m}{c^2 r_3}
+\frac{2\Omega}{D-1}\frac{2G_N m}{c^2 r_3}\, ,\label{13}\\
h_{\mu\mu}&=&-\frac{2}{D-1}\frac{2G_N m}{c^2 r_3} -2\left(\omega_{\mu}-\frac{\Omega}{D-1}\right)\frac{2G_N m}{c^2 r_3}\, ,\label{14} \ea
where $\Omega = \sum_{\mu=4}^D \omega_{\mu}$. Obviously, the second terms in Eqs. \rf{12}-\rf{14} are due to nonzero equations of state in the extra dimensions. {}From
\rf{14}, for the variation of the internal space volume we have
\be{15} \delta V_{D'} =\sum_{\mu=4}^D h_{\mu\mu}= -2\left(\frac{D-3}{D-1}+\frac{2\Omega}{D-1}\right)\frac{2G_N m}{c^2 r_3}\, , \ee
which shows that the variation arises due to two effects. The first term is due to the extra dimensions $(D>3)$, and the second term originates from nonzero equations
of state in the extra dimensions $(\Omega\neq 0)$. There is a very interesting special class of solutions, where both of these effects cancel each other:
\be{16}
\Omega=-\frac{D-3}{2}\, .
\ee
We call these solutions latent solitons. In this case, the internal space volume is constant:
\be{17} \delta V_{D'}=0\, , \ee
and $h_{00}$ and $h_{\alpha\alpha}$ exactly coincide with the Newtonian expressions and with each other: $h_{00}=\tilde h_{00}=h_{\alpha\alpha}=\tilde h_{\alpha\alpha}$.
Black stings $(D=4)$ and black branes $(D>4)$ are particular cases of the latent solitons with the same equations of state $\omega_{\mu} =-1/2$ in all extra dimensions.
For these particular cases, each $h_{\mu\mu}=0$, i.e. each scale factor of the internal spaces is constant. However, in the general case of the latent solitons,
to be in agreement with the observations it is sufficient to hold constant the total volume of the internal spaces. In this case the fifth force is absent. Since $\Omega
<0$, all or some of $\omega_{\mu}$ should be negative, i.e. such equations of state correspond to tension.

\section{Nonlinear models with toroidal compactification}

In this section we analyze nonlinear $f(R)$ KK models with toroidal compactification of the internal spaces. In the case of nonzero equations of state \rf{10} in the
extra dimensions, solutions (up to $O(1/c^2)$) read \cite{EZf(R)solitons}
\ba{18} h_{00}&=&-\frac{2(D-2)}{D-1}\frac{2G_N m}{c^2 r_3} -\frac{2\Omega}{D-1}\frac{2G_N m}{c^2 r_3}
-\frac{4b}{D-1}R\, ,\\
h_{\alpha\alpha}&=&-\frac{2}{D-1}\frac{2G_N m}{c^2 r_3} +\frac{2\Omega}{D-1}\frac{2G_N m}{c^2 r_3}
+\frac{4b}{D-1}R\, ,\label{19}\\
h_{\mu\mu}&=&-\frac{2}{D-1}\frac{2G_N m}{c^2 r_3} -2\left(\omega_{\mu}-\frac{\Omega}{D-1}\right)\frac{2G_N m}{c^2 r_3} +\frac{4b}{D-1}R\, ,\label{20} \ea
where $b\equiv (1/2)f''(0)$ and the scalar curvature is
\be{21}
R=\frac{1-\Omega}{2bD}\frac{2G_Nm}{c^2r_3}\exp\left[-\left(\frac{D-1}{4|b|D}\right)^{1/2}r_3\right]\, .
\ee
It is clear that the second terms in \rf{18}-\rf{20} take place due to the nonzero equations of state in the extra dimensions ($\omega_{\mu},\Omega \neq 0$) and the
third terms originate from the nonlinearity of the model $(b\neq 0)$. The Eq. \rf{21} shows that nonlinearity generates the Yukawa interaction with the mass
$[(D-1)/(4|b|D)]^{1/2}$ \cite{EZf(R)}.

\subsection{Pointlike mass\label{2.1s}}

Let us first consider the case of a pointlike matter source at rest, i.e. with the dustlike equation of state in all spatial dimensions: $\omega_{\mu} =0$,
$\mu=4,5,\ldots ,D$, $\Rightarrow$ $\Omega=0$. In this case, the second terms in Eqs. \rf{18}-\rf{20} disappear and we arrive at equations of the subsection \ref{1.1s}
with the admixture of the Yukawa terms. It can be easily seen that it is impossible to make constant the total volume of the internal spaces due to the Yukawa term in
the Eq. \rf{20}. In other words, the effect of nonlinearity can not reduce $\delta V_{D'}$ to zero. Similar to the linear case, the perturbations of the internal spaces
(the first term in the Eq. \rf{20}) result in the fifth force, which leads to the contradiction to the observational data \cite{EZf(R)}.

\subsection{Asymptotic latent solitons, black strings and black branes\label{2.2s}}

Now, we want to consider solutions, which are in agreement with the gravitational tests (the deflection of light and the time delay of radar echoes). In the case of the
linear models, it takes place for the latent solitons. That is we should take into account tension in the internal spaces: $\Omega \neq 0$. Unfortunately, it is
impossible to get exact soliton solutions in the case of an arbitrary function $f(R)$. Therefore, in the paper \cite{EZf(R)solitons}, it was proposed two types of
asymptotic solutions, where $h_{00}=\tilde h_{00}$ and $h_{\alpha\alpha}=\tilde h_{\alpha\alpha}\; \Rightarrow \; h_{00}=h_{\alpha\alpha}$. These asymptotic latent
solitons exist in the regions $r_3\gg\sqrt{|b|}$ and $r_3\ll\sqrt{|b|}$. Let us consider these two regions separately.

\subsubsection{$r_3\gg\sqrt{|b|}$}

In this asymptotic region, the exponent in \rf{21} is negligible and we can drop the third terms in Eqs. \rf{18}-\rf{20}. That is the effect of nonlinearity is
negligibly small and we arrive back at the case of the subsection \ref{1.2s}. Here, $\Omega =-(D-3)/2$ and $\delta V_{D'}=0$ in the Eq. \rf{15}, because the effects of
multidimensionality and tension cancel each other. Hence, the admixture of the fifth force is absent.

\subsubsection{$r_3\ll\sqrt{|b|}$}

In this case, we can replace the exponent in \rf{21} by unity. Here, the effect of nonlinearity is not negligible. After substitution \rf{21} into \rf{18} and \rf{19}
we get
\ba{22} h_{00}&=&-\frac{2G_N m}{c^2r_3}\left[1+\frac{D-3}{D-1}+\frac{2\Omega}{D-1}+\frac{2(1-\Omega)}{D(D-1)}\right]\, ,
\phantom{1+2}\\
h_{\alpha\alpha}&=&-\frac{2G_N m}{c^2r_3}\left[1-\frac{D-3}{D-1}-\frac{2\Omega}{D-1}-\frac{2(1-\Omega)}{D(D-1)}\right]\, , \label{23} \ea
where we have extracted from the first terms in (18) and (19) the Newtonian expression (the first terms in (22) and (23)). Obviously, the second terms in (22) and (23)
are due to the extra dimensions ($D>3$). The third and fourth terms appear due to the combined effect of tension in the extra dimensions and nonlinearity of the model.

It can be easily seen that for
\be{24}
\Omega = -\frac{D-2}{2}
\ee
the fifth force admixture is absent due to mutual annihilation of the effects of multidimensionality, tension and nonlinearity. For this value of $\Omega$, we have
$h_{00}=\tilde h_{00}$ and $h_{\alpha\alpha}=\tilde h_{\alpha\alpha}\; \Rightarrow \; h_{00}=h_{\alpha\alpha}$, and the considered model satisfies the gravitational
tests. An interesting feature of this case is that the value of the internal spaces total volume is not constant and its variation up to a sign coincides with the
Newtonian expression:
\be{25}
\delta V_{D'}=\sum_{\mu=4}^D h_{\mu\mu}=\frac{2G_N m}{c^2r_3}\, .
\ee
To understand the reason why we have agreement with the experiments for nonzero $\delta V_{D'}$, we  consider the sum \rf{25} in more detail:
\ba{26} \frac12\sum_{\mu=4}^D h_{\mu\mu} = \frac{2G_N m}{c^2r_3}\left[-\frac{D-3}{D-1} -\frac{6\Omega(D-1)+(D-2)^2+D+2}{2D(D-1)}+\frac12\right]\, , \ea
where the first term arises due to the extra dimensions (see \rf{9}) and the second term is the combined effect of tension and nonlinearity, from which we have singled
out 1/2. It can be easily verified that exactly for $\Omega =-(D-2)/2$, the combined effect of tension and nonlinearity eliminates the effect of multidimensionality.
Only in this case $\delta V_{D'} = -2\varphi_N/c^2$.

It can be easily seen that the value $\Omega =-(D-3)/3$ results in $\delta V_{D'}\equiv 0$. It takes place, e.g., for $\omega_{\mu}=-1/3\, , \mu =4,\ldots ,D$. For these
values of $\omega_{\mu}$, we have $h_{\mu\mu}=0$, that resembles the case of black strings and black branes. However, in contrast to the black strings/branes, here
$h_{00}\ne h_{\alpha\alpha}$. Therefore, we have shown that the demand $\delta V_{D'}= 0$ is neither necessary nor sufficient condition for satisfying the gravitational
tests. This very interesting result is certainly valid up to $O\left(1/c^2\right)$. It makes sense to check this assertion for the higher order of accuracy, e.g., up to
$O\left(1/c^4\right)$. We leave this problem for our further research.


\section{Spherical compactification of the internal space}

In this section we consider a model with spherical compactification of the internal space, where the background metrics is defined on a product manifold $M_4\times M_2$.
Here, $M_4$ describes the external four-dimensional flat spacetime and $M_2$ corresponds to a two-dimensional sphere with the radius (the internal space scale factor)
$a$. To create such spacetime with the curved internal space, we should introduce a background matter. As we have shown in our papers \cite{ChEZ1,ChEZ2}, this matter
simulates a perfect fluid with the vacuum-like equation of state in the external/our space. In the internal space (the two-sphere) the parameter of equation of state is
\be{27}
\omega_1=\frac{\Lambda_6}{1/[(2S_5\tilde G_6/c^4)a^2]-\Lambda_6}\, ,
\ee
where $\Lambda_6$ is a bare multidimensional cosmological constant. Different forms of matter can simulate such perfect fluid. For example, $\omega_1=1$ and $\omega_1=2$
correspond to the monopole form-fields (the Freund-Rubin scheme of compactification) and the Casimir effect, respectively. In the case $\Lambda_6 =0$ we get the
dustlike equation of state $\omega_1=0$. It is worth noting that for $\omega_1>0$ the internal space is stabilized \cite{ChEZ2}. In this model, the Eq. \rf{6} takes the
form $4\pi G_N = S_5 \tilde G_6/(4\pi a^2)$, where we take into account that the unperturbed volume of the internal space $V_{D'}\equiv V_2=4\pi a^2$. The background
metrics and matter are perturbed by a pointlike massive source with dustlike equations of state in all spatial dimensions. In the case $\omega_1>0$ solutions (up to
$O(1/c^2)$) read \cite{ChEZ2}
\ba{28}
h_{00} &=& -\frac{2G_N m}{c^2r_3} +\frac12\delta V_2\, ,\\
h_{\alpha\alpha} &=& -\frac{2G_N m}{c^2r_3} -\frac12\delta V_2\, ,\label{29}
\ea
where the relative value of the conformal variation of the volume of the two-sphere is
\be{30} \delta V_2 = -\frac{2G_N m}{c^2r_3}\exp\left(-\frac{\sqrt{\omega_1}}{a}r_3\right)\, . \ee
Therefore, this conformal variation generates the Yukawa interaction with the mass squared $\omega_1/a^2$. Obviously, the admixture of such interaction to $h_{00}$ and
$h_{\alpha\alpha}$ is negligible for sufficiently large Yukawa mass. Exactly this situation takes place for the gravitational tests in the Solar system \cite{ChEZ2}.
Here, $h_{00}=h_{\alpha\alpha}$ with very high accuracy, and we achieve a good agreement with the gravitational tests for the considered model.

Obviously, models with $\omega_1\leq 0$ do not satisfy the experimental data. For example, if $\omega_1 =0$, then Eqs. \rf{28} and \rf{29} exactly reduce to Eqs. \rf{7}
and \rf{8} for $D=5$ \cite{ChEZ1}. Here, the admixture of fifth force spoils the picture.

\section{Conclusion}

In this paper we investigated the cause of failure with the classical gravitational tests for models with toroidal compactification of the internal spaces. To perform
such investigations, we consider the weak-field limit of the corresponding KK models. Similar to general relativity, the matter source is taken in the form of a
pointlike mass at rest. In this case $T_{00}$ is the only nonzero component of the energy-momentum tensor. In the language of a perfect fluid, it means the dustlike
equation of state in all spatial dimensions. First, we singled out explicitly the contribution of the variations of the internal space volume to nonrelativistic
gravitational potentials. We have shown that these perturbations generate the fifth force. This is the reason of the problem. To avoid it, it is necessary to eliminate
the admixture of the fifth force to the nonrelativistic gravitational potential. It happens when the perturbation of the volume of the internal spaces is exactly equal
to zero: $\delta V_{D'}=0$. To achieve it, we should introduce tension in the internal spaces. For example, in the case of latent solitons (in particular, black strings
and black branes) we demonstrated how tension exactly eliminates the effect of extra dimensions, and the fifth force is absent.

In the case of nonlinear $f(R)$ models with toroidal compactification we again need tension to be in agreement with the observations. Here, we found two different types
of asymptotic latent solitons. One of them reproduces the results of the linear models, where tension cancels the effect of extra dimensions and the internal space
volume is constant. For the second type of soliton solutions, mutual annihilation of the effects of multidimensionality, tension and nonlinearity eliminates the fifth
force contribution. However, the variation of the internal space volume is not vanishing but coincides up to a sign with the Newtonian gravitational potential. Moreover,
for this type of models, we can make the variations of the internal space volume equal to zero: $\delta V_{D'}=0$, but then the gravitational tests fail here. This is a
very interesting example which indicates that the demand $\delta V_{D'}= 0$ is neither necessary nor sufficient condition for satisfying the gravitational tests.

It is well known that Kaluza-Klein models can be also formulated in the language of effective four-dimensional theories where the metric coefficients of the internal
spaces play a role of scalar fields/radions in the external spacetime. Both of these approaches are equivalent to each other. In appendix, we demonstrate this
equivalence by a simple example of a linear five-dimensional model with toroidal compactification.

As we noted in the introduction, the notion of tension follows from
the first law for black hole spacetimes \cite{TF,TZ,HO,TK}. However, the physical meaning of tension for ordinary astrophysical objects (such as our Sun) is still not
clear.

In the case of spherical compactification of the internal space, the situation is different from the cases discussed above. Here, instead of tension, we need a
background matter (e.g., monopole form-fields, the Casimir effect, etc.), which stabilizes the internal space. Then, the conformal perturbations of the internal space
generate the Yukawa interaction. For large Yukawa masses, the effect of this interaction is negligibly small, and the considered model satisfies the gravitational tests
at the same level of accuracy as general relativity. That is the case for our model in the Solar system.



\section*{Acknowledgements}

This work was supported in part by the "Cosmomicrophysics-2" programme of the Physics and Astronomy Division of the National Academy of Sciences of Ukraine. A.~Zh.
acknowledges the hospitality of the Theory Division of CERN where a part of this  work was carried out. We also thank Jennie Traschen for her useful comments.


\section*{Appendix: Effective four-dimensional picture}
\renewcommand{\theequation}{A\arabic{equation}}
\setcounter{equation}{0}

Above, we have considered the problematic aspects of Kaluza-Klein models in terms of multidimensional approach. It is known that these models can be also formulated in
the language of effective four-dimensional theories where the metric coefficients of the internal spaces play a role of scalar fields/radions in the external spacetime.
Both of these approaches are equivalent to each other. In this appendix we want to demonstrate this equivalence
by an example of a linear five-dimensional model investigated in Sec. 2. For this model, the Eq. \rf{2} with the help of dimensional reduction can be rewritten as the
following system of equations (see, e.g., \cite{OW}):
\ba{a1}
R_{ab}^{'(4)}&-&\frac{1}{2}R^{'(4)}g'_{ab}\approx\frac{8\pi G_N}{c^4}T_{ab}\, ,\\
\label{a2}
\triangle\sigma&\approx&-\frac{8\pi G_N}{c^4}\cdot\frac{2}{3}\left(T_A^A-\frac{3}{2}T_a^a\right)\, ,
\ea
where the prime denotes the quantities in the Einstein frame with conformally transformed metric coefficients $g'_{ab}=\exp(-\sigma)g_{ab}$. Throughout this appendix,
small Latin indices $a,b$ run over 0,1,2,3, Greek indices $\alpha,\beta$ run over 1,2,3 and capital Latin indices $A,B$ run over 0,1,2,3,4. This system of equations is
written in the weak field approximation. It means that the radion field $\sigma \sim h_{44}\sim O(1/c^2)$ (therefore, $\exp(-\sigma)\approx 1+O(1/c^2)$), and
$\triangle=\sum_{\alpha=1}^3\partial^2/\partial {x^{\alpha}}^2$.

Obviously, the Eq. \rf{a1} coincides with the Einstein equation in General Relativity. Therefore, for the spherically symmetric gravitating source the metric
coefficients are $g'_{00}\approx1+2\varphi_N/c^2$ and $g'_{\alpha\alpha}\approx-1+2\varphi_N/c^2$.

First, we consider the case of tension. In this case the only non-zero energy-momentum tensor components are $T_0^0\approx \rho c^2$ and $T_4^4\approx(1/2)\rho c^2$.
Then, $T_A^A\approx(3/2)\rho c^2$ and $T_a^a\approx\rho c^2$. Therefore, the right hand side of the Eq. \rf{a2} turns to zero and the solution is $\sigma\approx0$, where
we took into account the boundary conditions at the center and at infinity. Hence, in the case of tension (with the parameter of the equation of state $-1/2$) radion
does not contribute to the "true" metric components $g_{ab}$ and we have agreement with gravitational tests.

Second, we consider a delta-shaped matter source with dust-like equations of state in all spaces. Therefore, for the only non-zero component $T_0^0\approx \rho c^2$ we
have $T_A^A-(3/2)T_a^a\approx-(1/2)\rho c^2$, which results in the following solution of the Eq. \rf{a2}: $\sigma\approx(1/3)2\varphi_N/c^2$. Then the original (i.e. in
the Brans-Dicke frame) metric coefficients are
\ba{a3}
g_{00}&\approx& g'_{00}(1+\sigma)\approx1+\frac{4}{3}\frac{2\varphi_N}{c^2}\, ,\\
g_{\alpha\alpha}&\approx& g'_{\alpha\alpha}(1+\sigma)\approx-1+\frac{2}{3}\frac{2\varphi_N}{c^2}\, , \ea
in full analogy with Eqs. \rf{3} and \rf{4} for $D=4$. Obviously, this case without tension contradicts the observations.


\end{document}